\documentstyle[12pt,epsf,epsfig]{article}
\newcommand{\eg}{{\it e.g. }}
\newcommand{\mrm}[1]{\mbox{\rm #1}}
\newcommand{\beq}{\begin{equation}}
\newcommand{\eeq}{\end{equation}}
\newcommand{\bea}{\begin{eqnarray}}
\newcommand{\eea}{\end{eqnarray}}
\newcommand{\rfn}[1]{(\ref{#1})}

\newcommand{\nn}{\nonumber}

\def\lsim{\mathrel{\vcenter{\hbox{$<$}\nointerlineskip\hbox{$\sim$}}}}
\def\gsim{\mathrel{\vcenter{\hbox{$>$}\nointerlineskip\hbox{$\sim$}}}}

\def\ve{\epsilon}

\newcommand{\scr}{\scriptscriptstyle}

\newcommand{\snj}{\widetilde{N_j^c}}

\newcommand{\gnj}{\gamma_{\scr N_j}}
\newcommand{\gsnj}{\gamma_{\scr\snj}}

\newcommand{\Ynj}{Y_{\scr N_j}}

\newcommand{\Ynjeq}{Y_{\scr N_j}^{\rm eq}}

\newcommand{\Ynjratio}{{\Ynj\over\Ynjeq}}

\newcommand{\Yp}{Y_{j_+}}
\newcommand{\Ym}{Y_{j_-}}
\newcommand{\Ysnjeq}{Y_{\scr \snj}^{\rm eq}}
\newcommand{\Ypratio}{{\Yp\over\Ysnjeq}}

\newcommand{\Ymratio}{{\Ym\over\Ysnjeq}}

\newcommand{\YL}{Y_{\scr L_f}}
\newcommand{\YLt}{Y_{\scr L_s}}
\newcommand{\Yleq}{Y_l^{\rm eq}}
\newcommand{\Ylteq}{Y_{\wt{l}}^{\rm eq}}
\newcommand{\YLratio}{{\YL\over\Yleq}}
\newcommand{\YLtratio}{{\YLt\over\Ylteq}}

\def\wt#1{\widetilde{#1}}                    

\begin{document}
\begin{titlepage}
\pagestyle{empty}
\baselineskip=21pt
\rightline{CERN--TH/2002-139}     
\vskip 0.5in
\begin{center}
{\large{\bf 
Observable Consequences of Partially Degenerate Leptogenesis 
}}
\end{center}
\begin{center}
\vskip 0.25in
{
{\bf John Ellis}$^{1}$,
{\bf Martti Raidal}$^{1,2}$ and 
{\bf T. Yanagida}$^{1,3}$
\vskip 0.15in
{\it
$^1${Theory Division, CERN, CH-1211 Geneva 23, Switzerland}\\
$^2${National Institute of Chemical Physics and Biophysics, 
Tallinn 10143, Estonia}\\
$^3${Department of Physics, University of Tokyo, Tokyo 113, Japan}\\ 
}}
\vskip 0.65in
{\bf Abstract}
\end{center}
\baselineskip=18pt \noindent

In the context of the seesaw mechanism, it is natural that the large solar
and atmospheric neutrino mixing angles originate separately from large
$2\times 2$ mixings in the neutrino and charged-lepton sectors,
respectively, and large mixing in the neutrino couplings is in turn more
plausible if two of the heavy singlet neutrinos are nearly degenerate. We
study the phenomenology of this scenario, calculating leptogenesis by
solving numerically the set of coupled Boltzmann equations for
out-of-equilibrium heavy singlet neutrino decays in the minimal
supersymmetric seesaw model.  The near-degenerate neutrinos may weigh
$\lsim 10^8$~GeV, avoiding the cosmological gravitino problem.  This
scenario predicts that $Br(\mu\to e\gamma)$ should be strongly suppressed,
because of the small singlet neutrino masses, whilst $Br(\tau\to
\mu\gamma)$ may be large enough to be observable in B-factory or LHC
experiments. If the light neutrino masses are hierarchical, we predict
that the neutrinoless double-$\beta$ decay parameter $m_{ee} \approx
\sqrt{\Delta m_{sol}^2}\sin^2\theta_{12}$.

\vfill
\vskip 0.15in
\leftline{CERN--TH/2002-139}
\leftline{June 2002}
\end{titlepage}
\baselineskip=18pt


Neutrino oscillation data~\cite{skatm,sno1} are converging towards
unique solutions for both the solar and atmospheric neutrino
anomalies~\cite{sno2}. There are two large, almost maximal, mixing angles
$\theta_{12}$ and $\theta_{23}$ in the light neutrino mass matrix, that
give rise to the solar and atmospheric neutrino oscillations,
respectively, whilst the third mixing angle $\theta_{13}$ is constrained
to be small~\cite{CHOOZ,Boehm}. This pattern motivates theoretical
approaches based on two $2\times 2$ mixings instead of one general
$3\times 3$ mixing.

The smallness of the neutrino masses is generally explained via the seesaw
mechanism~\cite{seesaw}, described by the superpotential~\footnote{We 
assume low-energy supersymmetry, which leaves unaltered the flavour 
parameters, while providing extra low-energy observables.}: 
\begin{eqnarray}
\label{w}
W = N^{c}_i (Y_\nu)_{ij} L_j H_2
  -  E^{c}_i (Y_e)_{ij}  L_j H_1 
  + \frac{1}{2}{N^c}_i (M_N)_{ij} N^c_j + \mu H_2 H_1 \,.
\label{suppot}
\end{eqnarray}
Here the indices $i,j$ run over three generations and $M_N$
is the heavy singlet-neutrino mass matrix. We shall work in a basis where 
$(M_N)_{ij}$ is real and diagonal, $(M_N)_{ij} = { M_{N_i}} \delta_{ij},$
and we define $M_{N_1}<M_{N_2}<M_{N_3}.$
In analogy with the CKM mixing matrix in the quark sector, 
the mixing matrix $V_{MNS}$ measured in neutrino oscillations
is a product of two matrices $V_{MNS}=U_{e}^\dagger U_\nu,$ where
$U_{\nu}$ diagonalizes the light-neutrino seesaw mass matrix ${\cal 
M}_\nu$
\bea 
{\cal M}_\nu={Y}_\nu^T \left({ M_N}\right)^{-1} 
{Y}_\nu v^2 \sin^2\beta ,
\label{seesaw1}
\eea
according to
\bea
U_{\nu}^T {\cal M}_\nu U_{\nu} = {\cal M}^D_\nu\,,
\label{Mnud}
\eea
and $U_e$ helps diagonalize the charged-lepton Yukawa coupling matrix 
$Y_e$ in (\ref{suppot}):
\bea
V_e^\dagger Y_e U_e=Y_e^D .
\label{Yed}
\eea
Deriving the observed neutrino mixing angles from the neutrino Dirac
Yukawa couplings $(Y_\nu)_{ij}$ and Majorana masses $M_{N_i}$ is
problematic. A generic difficulty, in view of the hierarchy $\Delta
m^2_{sol}\ll\Delta m^2_{atm},$ is that large mixing angles for both solar
and atmospheric mixings can be obtained only at the price of some fine
tuning. Technically speaking, the (23) sub-determinant of the
light-neutrino mass matrix must vanish.  While Yukawa textures which may
accommodate this feature can be obtained from well-motivated physics
ideas~\cite{gn} such as GUTs, Abelian and non-Abelian flavour symmetries,
democratic principles, etc., many of these approaches feature unknown
model coefficients of order unity that cannot be predicted without further
assumptions, failing which they must be tuned {\it a posteriori}.

One way to attack this problem is to note that the observed leptonic
mixing pattern can be regarded as the combination of two $2\times 2$
mixings in the neutrino and the charged-lepton Yukawa couplings
$(Y_\nu)_{ij}$ and $(Y_e)_{ij},$ respectively. If the matrices $U_{\nu}$
and $U_{e}$ contain non-trivial mixings only in the (12) and (23)
sub-matrices, respectively, large mixing angles $\theta_{12}$,
$\theta_{23}$ and vanishing $\theta_{13}$ follow naturally.  Notice,
however, that the matrices $U_{\nu}$ and $U_{e}$ are of different nature.
Whilst $U_{e}$ directly rotates the superpotential couplings in
(\ref{suppot}), $U_{\nu}$ diagonalizes the effective mass matrix obtained
from the superpotential couplings via the seesaw relation (\ref{seesaw1}).
Therefore, one may expect some differences between these two matrices
$U_{\nu}$ and $U_{e}$, which may be the reason why the solar mixing
somewhat differs from the atmospheric one~\footnote{The above breakdown is
basis-dependent: one could choose to work in the basis in which $Y_e^D$ is
diagonal, in which case the mixing in $U_{e}$ simply moves into $Y_{\nu}$,
while the physics remains unchanged.}.

Such a structure for  $Y_{\nu}$ and $U_\nu$
would arise naturally if there is an (approximate) $S_2$ symmetry 
between the first and the second
generation fields: $N_1\leftrightarrow N_2$ and $L_1\leftrightarrow L_2,$
implying
\begin{equation}
Y_\nu =
\left(\begin{array}{cc}
 1 & 1 \\
 1 & 1 
\end{array}\right)\, , \;\;\; 
M_N = 
\left(\begin{array}{cc}
 1 & 0 \\
 0 & 1 
\end{array}\right)\, ,
\label{ansatz}
\end{equation}
in the (12) sector.
The large mixing angle $\theta_{12}$ diagonalizing $Y_{\nu}$ follows 
immediately from this approximate symmetry, as does the prediction
that $M_{N_1}\sim M_{N_2},$ , while $M_{N_3}$ may be different. 
Larger flavour symmetries would be needed to extend this approach to 
the third generation, to incorporate the $E^c_i$ fields, and to explain 
the 
largeness of $\theta_{23}.$ We do not pursue such model-building issues
here, but rather pursue the phenomenological implications of
the proposed structure (\ref{ansatz}) for $Y_{\nu},$ $Y_{e}$ and $M_{N}$, 
namely that
the Dirac Yukawa matrices $Y_{\nu}$ and $Y_{e}$ have non-trivial 
$2\times2$ (12) and
(23) sub-matrices and non-vanishing diagonal elements (33) and (11),
respectively, whilst the rest of their entries vanish and $M_{N}$ has 
a pair of (almost) degenerate eigenvalues. We incorporate the low-energy 
neutrino data into our parametrization of the Yukawa couplings, 
and systematically scan over all the remaining
free parameters of the seesaw model. In this way we include
theoretical models of~\cite{gn} that predict similar patterns.

We focus our attention on leptogenesis~\cite{fy} in this framework, 
finding that the pair of near-degenerate heavy singlet neutrinos enable one to 
lower the scale of thermal leptogenesis below the gravitino
bound on the reheating temperature of the Universe in
supergravity~\cite{gravitino} and gauge-mediated 
models~\cite{gmgravitino}. We study implications of this scenario for
$\beta\beta_{0\nu}$ decay~\cite{bbth} and on charged-lepton
flavour-violating (LFV) decays~\cite{bm}. A important aspect of this
scenario is that there are less free parameters than in the general seesaw
model, implying testable phenomenological consequences.  We find that
$\tau \to \mu \gamma$ may be observable, whilst $\mu \to e \gamma$ is
suppressed, and make a specific prediction for neutrinoless double-$\beta$
decay.

We start by counting the physical parameters of the model. In the chosen
basis the Yukawa matrix $Y_{\nu}$ contains five complex parameters,
and the basis for $N_i$ is completely fixed. Since three phases can be 
removed by redefining the $L_i$ fields, only two phases are physical.
One can parametrize $Y_{\nu}$ as
\bea
(Y_\nu)_{ij} = Z^\star_{ik} {Y}^D_{\nu_k} X^\dagger_{kj},
\label{Y}
\eea
where the unitary matrices $X$ and $Z$ contain only (12) mixing, and hence
only one mixing angle. The matrix $X$ is real, while the matrix $Z$ may be
written as $Z \equiv P_1 \overline{Z} P_2,$ where $\overline{Z}$ is a real
matrix. In this `high-energy' basis, the diagonal matrices $P_{1,2} \equiv
\mrm{diag}(e^{i\theta_{1,2}}, 1, 1 )$ contain the two physical 
CP-violating phases. Diagonalizing $Y_{e}$
according to (\ref{Yed}), the real parameters correspond to three diagonal
Yukawa couplings $Y_{e}^D$ and one mixing angle in each of the rotation
matrices $U_{e}$ and $V_{e}.$ Again, three out of five phases
can be absorbed into redefinition  of right-handed fields $E^c_i.$
Thus the basis for $E^c_i$ is now completely fixed and there is one 
physical phase in the mixing matrix $U_{e}=\overline{U_e}P_3,$ 
where $\overline{U_e}$ is real and 
$P_{3} \equiv\mrm{diag}(1, 1, e^{i\theta_{3}} )$. 
The mixing and one phase in $V_{e}$ are unobservable.
This implies that, in the basis in which the charged lepton masses are 
diagonal,  we have a total of 9 physical
parameters in the neutrino Yukawa couplings, which together with the 3 unknown
heavy masses $M_{N_i}$ make a total of 12 parameters in this version of
the minimal seesaw model. This should be compared with the 18 physical
parameters in the general case, implying more predictivity and hence
better possibilities to test the scenario at high and low energies.

We recall there are three types of leptonic observables in supersymmetric
seesaw models: (i) leptogenesis via out-of-equilibrium decays of heavy
singlet neutrinos and (ii) renormalization-induced mixings in the slepton
mass matrix due to the off-diagonal $Y_\nu$ couplings, as well as (iii)
the light neutrino masses and mixings. The leptogenesis CP asymmetry
produced in the out-of-equilibrium decays of each of the $N_i$ is given
by~\cite{p,k-sm,k-mssm}
\begin{eqnarray}
\epsilon_i &=& -\frac{1}{8 \pi} \sum_{l} 
\frac{ \mbox{Im}\Big[
\left( { Y_\nu}{ Y_\nu}^\dagger  \right)^{il}
\left( { Y_\nu}{ Y_\nu}^\dagger \right)^{il}
\Big]}
{ \sum_{j} |{ Y_\nu}^{ij}|^2 }
\nn \\
& &
\sqrt{x_l} \Big[  \mbox{Log} (1+1/x_l) +  \frac{2}{(x_l-1)}\Big] , 
\label{eps}
\end{eqnarray}
where $x_l \equiv (M_{N_l} / M_{N_i})^2.$ It is clear from (\ref{eps})
that the generated asymmetry depends only on 
\begin {equation}
\label{yy+1}
Y_\nu Y_\nu^\dagger 
 = P_1^\star \overline{Z}^\star (Y_\nu^D)^2 \overline{Z}^T P_1\, 
\end{equation}
and on the heavy neutrino masses. Notice that $Y_\nu Y_\nu^\dagger$ 
is independent of the left-rotations of $L_i$, and is therefore
independent of the $Y_e$ basis. 
Therefore, in our scenario the baryon
asymmetry of the Universe depends only on the single high-energy CP phase
in $P_1.$ If the CP-conserving parameters are known, this can be
calculated from the observed baryon asymmetry.

Renormalization of soft supersymmetry-breaking parameters due to the
presence of $Y_\nu$ above the heavy-neutrino decoupling scales modifies
the left-slepton mass matrix $m_{\tilde L}$ and trilinear soft
supersymmetry-breaking $A_e$ terms. In the basis of diagonal
charged leptons one has in leading-logarithmic order~\cite{bm}
\begin{eqnarray}
(\delta m_{\tilde{L}}^2)_{ij}&\simeq&
-\frac{1}{8\pi^2}(3m_0^2+A_0^2) (Y^\dagger L Y)_{ij} \,,
\nonumber\\
(\delta A_e)_{ij} &\simeq&
-\frac{1}{8\pi^2} A_0 Y_{e_i} (Y^\dagger L Y)_{ij} \,,
\label{leading}
\end{eqnarray}
which are proportional to
\begin{equation}
Y^\dagger L Y = 
P_3^* {\overline U_e}^\dagger X Y^D P_2 {\overline Z}^T L {\overline Z}^*
P_2^* Y^D X^\dagger  {\overline U_e} P_3,
\label{YYren}
\end{equation}
where $L$ is a diagonal matrix: $L_{ij}=\ln (M_{GUT}/M_{N_i})  
\delta_{ij}.$ Note that the high-energy CP-violating phases in
(\ref{YYren}) are those in $P_2$ and $P_3.$ The CP-violating observables 
in LFV
processes all depend on a single CP-violating invariant $J_\nu={\rm Im}
H_{12} H_{23} H_{31}$~\cite{lfvcp}, where $H=Y_\nu^\dagger L Y_\nu$. This
influences slepton physics at colliders and also determines the T-odd
asymmetry in $\mu\rightarrow 3e$~\cite{3l}. Therefore, all low-energy
CP-violating observables measure one combination of phases in $P_{2,3}.$ 
Due to the near-degeneracy of the heavy neutrino masses, the
lepton EDMs are unobservably small in this scenario~\cite{ehrs2}.

So far, we have considered the high-energy parametrization of the neutrino
sector in terms of the diagonal Yukawa couplings and the unitary matrixes
$X$, $Z$, $U_e.$ Counting physical degrees of freedom is straightforward
in this approach, but it has limited applicability to low-energy
phenomenology. Therefore we develop a complementary `low-energy'
parametrization.  We start by discussing the light neutrino mass matrix
(\ref{seesaw1}) in our scheme. Of course, at
low energy it is most convenient to work in the basis in which
charged leptons are diagonal. 
After redefinitions of the $L_i$ fields, the diagonalizing matrix for
light neutrinos becomes
\bea
U_\nu=V_\nu P_0,
\eea
where $V_\nu$ is real and  contains the mixing angles $\theta_{12},$ 
$\theta_{23},$ and $P_0 \equiv
\mrm{diag}(1,e^{i\phi_2}, e^{i\phi_3} )$ contains two
Majorana phases $\phi_{2,3},$ whilst the third angle $\theta_{13}$ 
vanishes and the neutrino oscillation phase $\delta$ is absent.
Therefore, all the low-energy neutrino
observables, such as neutrino oscillations, $\beta\beta_{0\nu}$ decay,
etc., depend on the 7 effective low-energy parameters, which are functions
of the 12 parameters in (\ref{suppot}). We recall that, whilst neutrino
oscillations measure the mass-squared differences of neutrinos and their
mixing angles, $\beta\beta_{0\nu}$ decay measures one particular
combination of their masses and mixing matrix elements:
\bea
|m_{ee}| \equiv \left|\sum_i (U_\nu)_{ei}^* m_{\nu_i} 
(U_\nu)_{ie}^\dagger \right|=
\left| m_{\nu_1} \cos^2\theta_{12}  + 
m_{\nu_2} \sin^2\theta_{12}e^{2i\phi_2}  \right|.
\label{mee} 
\eea
The effect of the Majorana phase $\phi_2$ becomes visible only if
$m_{\nu_1}\sim m_{\nu_2}$, but its measurement is not straightforward even
in this case~\cite{bbth}. In the case of hierarchical neutrino masses 
$m_{\nu_1}\ll m_{\nu_2}$, we
have the definite prediction $m_{ee}= m_{\nu_2} \sin^2\theta_{12}.$

In order to satisfy automatically the oscillation data, the light neutrino
parameters must be an input of the parametrization. We therefore
define~\cite{ci}:
\bea { Y_\nu}= 
\frac{\sqrt{{M_N}} R \sqrt{{\cal M}^D_\nu}\, U_\nu^\dagger}{v\sin\beta},
\label{Ynu}
\eea
where $R$ is a complex orthogonal matrix. In our case, it has a
non-trivial $2\times 2$ (12) submatrix only, and is parametrized by just
one complex number. Examining the combination $Y_\nu Y_\nu^\dagger$ using
(\ref{Ynu}), we observe that the single complex phase in $R$ is
responsible for leptogenesis. On the other hand, (\ref{leading}) tells
that in this parametrization the renormalization-induced low-energy CP
violation depends also on the Majorana phases  in a non-trivial
way. One can exploit this to take a complete bottom-up approach, since in 
the supersymmetric seesaw model the low-energy degrees of freedom may in
principle be used to reconstruct all the high-energy neutrino
parameters~\cite{di}.  A parametrization related to the solutions of the
renormalization-group equations (RGEs) for the soft supersymmetry-breaking
slepton masses (\ref{leading}) was worked out in~\cite{ehrs}.

Our central objective in this Letter is to study the phenomenology of
leptons in the proposed scheme. In particular, we focus our attention on
leptogenesis~\footnote{Implications of particular neutrino mass textures
on leptogenesis have been studied in~\cite{k-sm,k-mssm,leptex,Branco}.} 
and its
relations to neutrino masses, $\beta\beta_{0\nu}$ decay and LFV
observables. A strong motivation is provided by the gravitino problem in
generic supergravity theories, which restricts the maximum reheating
temperature $T_R$ of the Universe after inflation. In
supergravity models with a gravitino mass of order 500 
GeV~\cite{gravitino} or in some gauge-mediated models~\cite{gmgravitino} 
the bound is 
\bea
T_R\lsim 10^8\;\;\mrm{GeV.}
\label{reh}
\eea 
On the other hand, if the baryon asymmetry of the Universe is due to the
decays of the lightest singlet neutrino $N_1,$ there is an
$M_{N_1}$-dependent upper bound on the CP asymmetry $\epsilon_1$
\cite{di2}. In the context of thermal leptogenesis, the observed baryon
asymmetry implies a lower bound $M_{N_1}\gsim 10^{10}$ GeV~\cite{er} which
is potentially in serious conflict with (\ref{reh}).

This problem could be overcame by abandoning thermal leptogenesis and
considering non-thermally produced neutrinos~\cite{nonth}, but such
scenarios are still speculative and lack predictivity. Another well-known
solution to the problem in the context of thermal leptogenesis is to
consider the decays of two heavy neutrinos which are approximately
degenerate in mass~\cite{p}. Because the self-energy contribution to
$\epsilon_i$ is enhanced for degenerate heavy neutrino masses, the
observed baryon asymmetry can be generated by moderately degenerate heavy
neutrinos that are relatively light, which is the option pursued here.
Since, in the proposed scenario, the $Y_\nu$ couplings spanning the (12)
submatrix are separate and distinct from the (33) element, it is natural
that the masses of the heavy neutrinos $N_{1,2}$ differ from that of
$N_{3}$~\footnote{We assume here that $N_{1,2}$ are not exactly degenerate
in mass, presumably because the (unspecified) underlying symmetry
principle for first two generations, such as the $S_2$ example mentioned
earlier, is weakly broken.}.

In the minimal supersymmetric seesaw model, the baryon asymmetry
originates from out-of-equilibrium decays of neutrinos and sneutrinos.  
Assuming $M_{N_3}> M_{N_2}\approx M_{N_1}$, we need consider only the two
lighter neutrino decays. Even if
$M_{N_2}\approx M_{N_1}$, the CP asymmetries $\epsilon_{1,2}$ may be very
different in magnitude. Although the products of the numerator and mass 
factors in 
(\ref{eps}) are identical for $\epsilon_{1,2}$, the denominators
$(Y_\nu Y_\nu^\dagger)_{11}$ and $(Y_\nu Y_\nu^\dagger)_{22}$ may be very 
different~\footnote{This would require the two-generation 
symmetry to be broken in the Yukawa couplings.}. Therefore,
$N_{1,2}$ may contribute to the lepton asymmetry and to the reaction rates
of the processes involved in very different ways, making the washout
processes non-trivial. The dominant processes which determine the order of
magnitude of the asymmetry are the $\Delta L=1$ neutrino and sneutrino
decays. The $\Delta L=2$ scatterings are suppressed by additional powers
of small Yukawa couplings, and are completely negligible in the
low-$M_{N_1}$ regime we consider here~\cite{k-sm}.  There are also $\Delta
L=1$ two-to-two scatterings involving top quarks and squarks. Those have
additional powers of $y_t^2/(4\pi)$ and make a contribution of relative
order unity to the washout processes: we neglect them for simplicity. In
supersymmetric models there are also processes transforming leptons into
scalar leptons and vice versa, e.g., $e+e\leftrightarrow\wt{e}+\wt{e},$
which we do take into account. 

In this approximation, defining $Y_{j\pm}
\equiv Y_{\scr\snj}\pm Y_{\scr\snj^{\dagger}}$ for the scalar neutrinos 
and
their antiparticles, the Boltzmann equations for the ratios of particle
densities divided by the entropy density, and for the lepton asymmetries
$\YL$ and $\YLt$ in fermions and scalars, respectively, are given
by~\cite{k-mssm}:
\bea
{\mbox{d}\Ynj\over\mbox{d}z}&=&{-z\over sH(M_{N_1})}
        \left(\Ynjratio-1\right) \gnj  \;,  \nn\\
      {{\mbox{d}\Yp\over\mbox{d}z}}&=&{-z\over sH(M_{N_1})}
        \left(\Ypratio-2\right)\gsnj \;,  \nn\\
      {\mbox{d}\Ym\over\mbox{d}z}&=&{-z\over sH(M_{N_1})}\left(
        \Ymratio
       -{1\over2}\YLtratio  +{1\over2}\YLratio \right) \gsnj \nn\;,\\
     {\mbox{d}\YL\over\mbox{d}z}&=&{-z\over sH(M_{N_1})}\left\{
        \sum\limits_j\left[\left({1\over2}\YLratio+\ve_j\right)
        \left({1\over2}\gnj+\gsnj\right)-{1\over2}\ve_j\left(
        \Ynjratio\gnj+\Ypratio\gsnj\right)
        \right.\right.  \nn \\
      &&\left.\left.+{1\over2}\Ymratio\gsnj\right] +\left(\YLratio
        -\YLtratio\right)\gamma_{\mbox{\tiny MSSM}}   \right\}\;,  \nn\\
      {\mbox{d}\YLt\over\mbox{d}z}&=&{-z\over sH(M_{N_1})}\left\{
        \sum\limits_j\left[\left({1\over2}\YLtratio+\ve_j\right)
        \left({1\over2}\gnj+\gsnj\right)-{1\over2}\ve_j\left(
        \Ynjratio\gnj+\Ypratio\gsnj\right)\right.\right. \nn \\
      &&\left.\left. -{1\over2}\Ymratio\gsnj\right]
     +\left(\YLtratio-\YLratio\right)\gamma_{\mbox{\tiny MSSM}}
\right\} \;,
\label{boltz}
\eea
where the temperature dependence is via $z=M_{N_1}/T,$ 
$Y^{\rm eq}_x$ denotes the equilibrium density of the particle denoted by
the subscript $x$. The reaction densities for neutrinos and
sneutrinos are given by~\cite{k-mssm}   
    \beq
      \gnj=2\,\gsnj = {M_{N_1}^4\over4\pi^3}\;
{\left(Y_\nu Y_\nu^\dagger\right)_{jj}}\;
        {a_j\sqrt{a_j}\over z}\mbox{K}_1(z\sqrt{a_j})\;,
    \eeq
where $a_j=(M_{N_j}/M_{N_1})^2,$ and that for the MSSM processes is
    \beq
       \gamma_{\mbox{\tiny MSSM}}\approx 
{M_{N_1}^4\,\alpha^2\over 4\pi^3}\,{1\over z^4}
       \left[\ln\left({4M_{N_1}^2\over z^2 m^2_{\wt{\gamma}}}\right)
       -2\gamma_{\mbox{\tiny E}}-3\right]\;,
    \eeq
where $m^2_{\wt{\gamma}}$ is the photino mass which contributes to
$e+e\leftrightarrow\wt{e}+\wt{e}$ in the $t$ channel. The total lepton 
asymmetry
$Y_L=\YL+\YLt$ is then converted by sphalerons into a baryon asymmetry:
$Y_B= C\,Y_L$, where $C=-8/15$ in the MSSM.

In our subsequent numerical analysis, we fix the known light neutrino
parameters to be $\Delta m^2_{32}=3\times 10^{-3}$ eV$^2,$ $\Delta
m^2_{21}=5\times 10^{-5}$ eV$^2,$ $\tan^2\theta_{23}=1$ and
$\tan^2\theta_{12}=0.4$, corresponding to the LMA solution for the solar
neutrino anomaly.  The angle $\theta_{13}$ is predicted to vanish,
together with the neutrino oscillation phase $\delta,$ 
and the connection between leptogenesis
and the oscillation phase proposed in~\cite{Branco} does not hold. We
assume the normal mass ordering for light neutrinos,
$m_{\nu_1}<m_{\nu_2}<m_{\nu_3},$ and generate $Y_\nu$
according to (\ref{Ynu}). All the unknown input parameters are generated
randomly as follows. The lightest light neutrino mass $m_{\nu_1}$ is
generated in the range $(10^{-5}-1)$ eV, the Majorana phases $\phi_{2,3}$ in
the range $(0-2\pi),$ the lightest heavy neutrino mass $M_{N_1}$ in the
range $(10^{6}$ - just above $10^{8})$ GeV, the degeneracy parameter
$\Delta \equiv \ln(M_{N_2}/M_{N_1}-1)$ in the range $(-15 -1),$ the 
heaviest
neutrino mass $M_{N_3}$ in the range $(2\times M_{N_2} -10^{15})$ GeV and
the complex parameter $r_{12}$ of the orthogonal matrix $R$ in the range
$|r_{12}|=(0-10)$ with an arbitrary phase. Each mass parameter is
generated with a flat distribution on a logarithmic scale. We solve the
Boltzmann equations \rfn{boltz} numerically, assuming initial thermal
abundances for neutrinos and sneutrinos, and require that the induced
baryon asymmetry be in the range~\cite{olive} 
\bea
3\times 10^{-11} \lsim Y_B \lsim 9\times 10^{-11}. 
\label{YB}
\eea
We require that the neutrino decay width satisfy
$\Gamma_{N_i} < 10^{-5} (M_{N_2}-M_{N_1})$, so that we are far from the
resonant region and can trust our perturbative calculation. Subsequently,
we study correlations between the leptogenesis parameters and the light
and heavy neutrino masses, the $\beta\beta_{\nu0}$ decay parameter
$m_{ee},$ and the LFV decays of the charged leptons.

\begin{figure}[htbp]
\centerline{\epsfxsize = 0.5\textwidth \epsffile{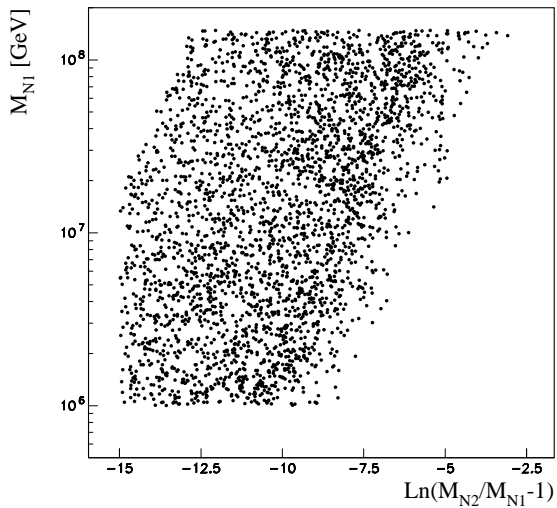} 
}
\caption{\it 
Scatter plot of the lightest singlet neutrino mass $M_{N_1}$ as a 
function  of the degeneracy parameter $\ln(M_{N_2}/M_{N_1}-1).$
The baryon asymmetry is required to be in the range (\ref{YB}).
\vspace*{0.5cm}}
\label{fig0}
\end{figure}

\begin{figure}[htbp]
\centerline{\epsfxsize = 0.5\textwidth \epsffile{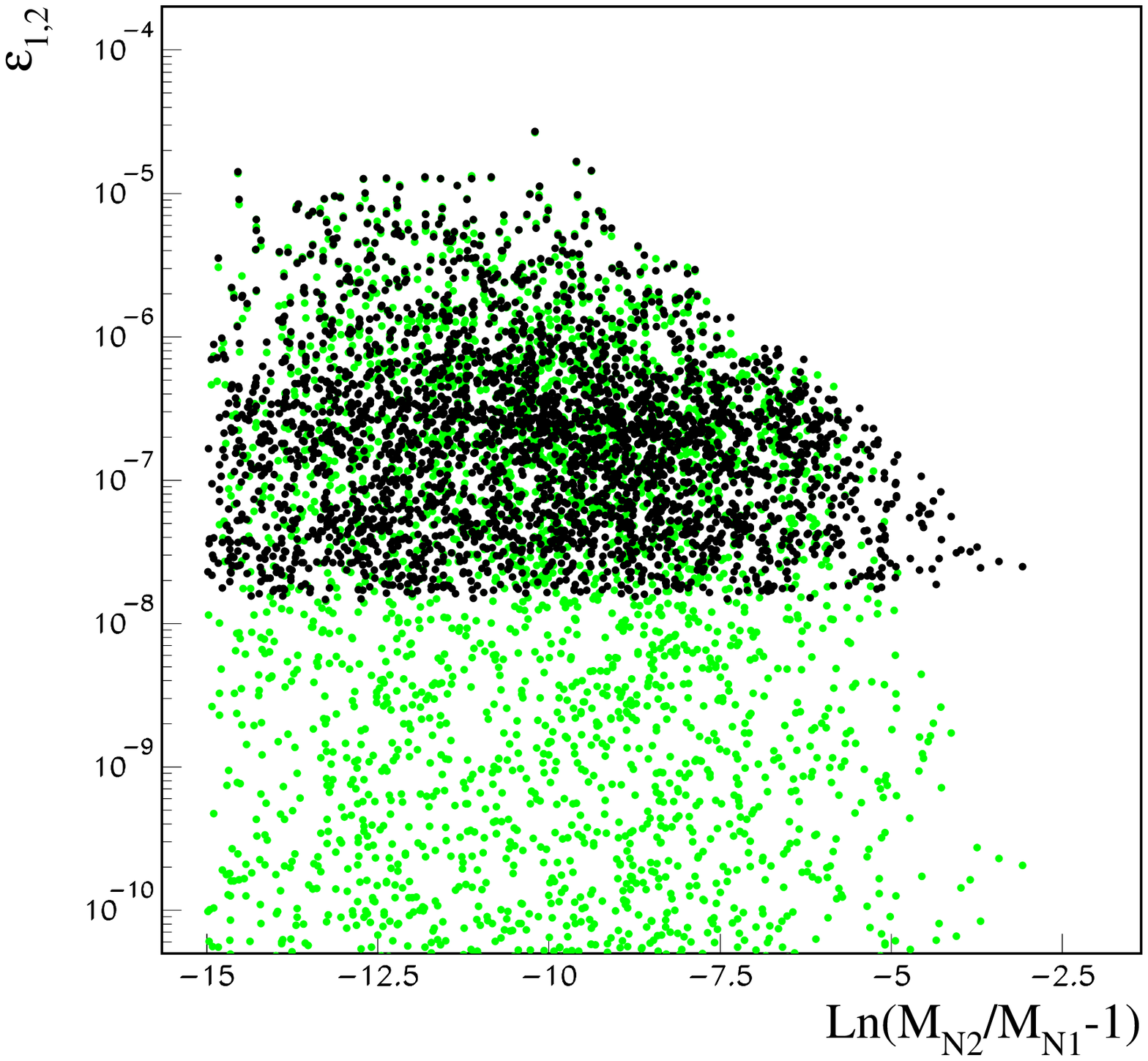} 
\hfill \epsfxsize = 0.5\textwidth \epsffile{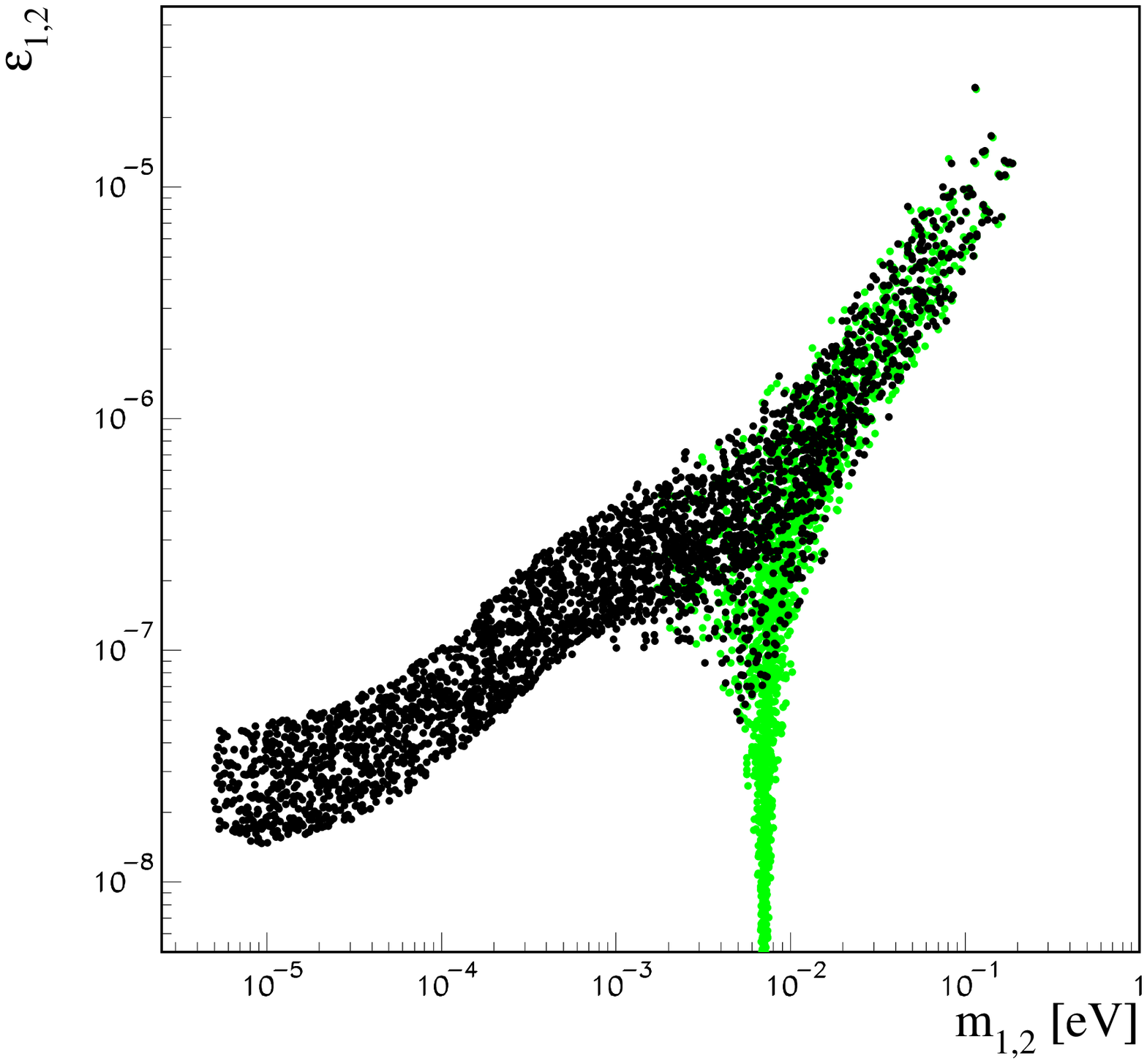} 
}
\caption{\it 
Scatter plots of the CP-violating asymmetries $\epsilon_{1,2}$
denoted by black and green (grey), respectively, as functions  of the 
degeneracy parameter $\ln(M_{N_2}/M_{N_1}-1)$ and the effective mass parameters
$\tilde m_{1,2}$. The baryon asymmetry is required to be in the range 
(\ref{YB}).
\vspace*{0.5cm}}
\label{fig1}
\end{figure}

We find a {\it weak} correlation of the allowed range of $M_{N_1}$ 
with $\Delta = \ln(M_{N_2}/M_{N_1}-1),$ as seen in Fig.~\ref{fig0}. 
We see
that successful leptogenesis is possible with light $N_{1,2}$ already if
they are degenerate in mass at the level of a few percent. 
Motivated by the gravitino problem, we
limit $M_{N_1}$ to the range $10^6$~GeV to about $10^8$~GeV in our 
subsequent numerical
examples. We find a {\it stronger} correlation between the leptogenesis 
parameters $\varepsilon_{1,2}$ and the
neutrino degeneracy parameter $\Delta.$ In Fig.~\ref{fig1} (a) we plot the
CP asymmetries $\epsilon_{1,2},$ denoted by black and green points,
respectively, as functions of the degeneracy parameter
$\Delta=\ln(M_{N_2}/M_{N_1}-1).$ There is a $\Delta$-dependent upper bound
on $\epsilon_{1,2},$ coming from the mass function in \rfn{eps}. 
Solutions to the Boltzmann equations
\rfn{boltz} are often discussed in the literature in terms of effective 
mass parameters
\bea
\tilde m_i \equiv \left(Y_\nu Y_\nu^\dagger \right)_{ii}
\frac{v^2 \sin^2\beta}{M_{N_i}}.
\label{tildem}
\eea
If leptogenesis originates from $N_1$ decays, as is the case for
hierarchical heavy neutrinos, fixing $Y_B$ implies almost a one-to-one
correspondence between $\epsilon_{1}$ and $\tilde m_1$ in the low
$M_{N_1}$ region~\cite{k-sm,er}. However, the solutions to \rfn{boltz} are
non-trivial, as seen in Fig.~\ref{fig1} (b), where we plot
$\epsilon_{1,2}$ versus the parameters $\tilde m_{1,2},$ respectively.  A
direct correlation is observed only for high values of $\tilde m_{1}\sim
\tilde m_{2}\gsim 5\times 10^{-2}$~eV, where both neutrinos $N_{1,2}$
contribute to $Y_B$ in a similar way. This region corresponds to large
and quasi-degenerate
light neutrino masses. Because of strong washout effects, $\epsilon_{1,2}$
must be large and, correspondingly, $M_{N_2}/M_{N_1}-1$ small in this
region.  For lower values of $\tilde m_{1}$ the non-linear shape of
Fig.~\ref{fig1} (b) indicates the non-trivial washout effects of $N_2.$

\begin{figure}[htbp]
\centerline{\epsfxsize = 0.5\textwidth \epsffile{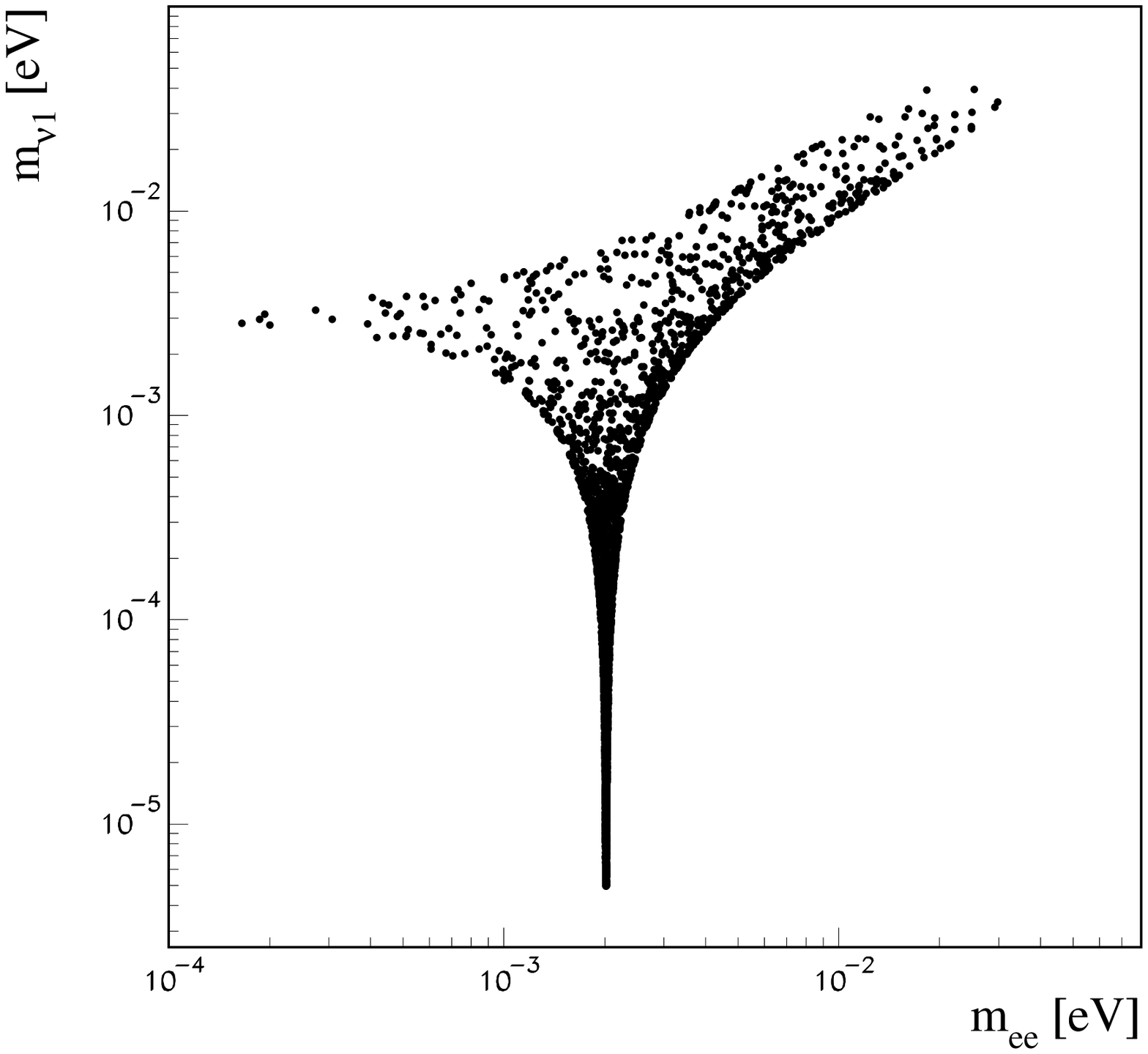} 
\hfill \epsfxsize = 0.5\textwidth \epsffile{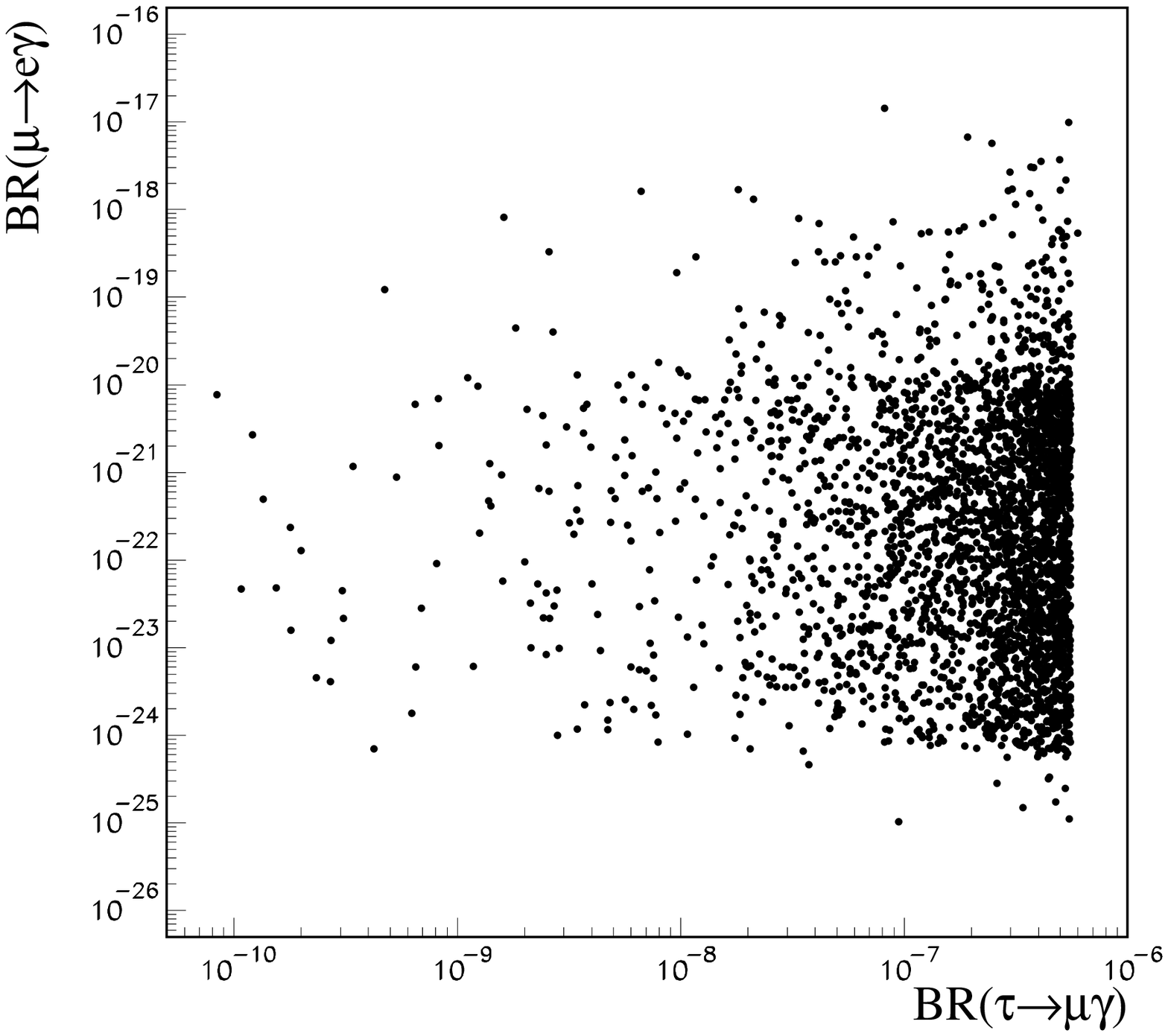} 
}
\caption{\it 
Scatter plots of the lightest neutrino mass $m_{\nu_1}$ versus the
$\beta\beta_{0\nu}$-decay parameter $m_{ee},$ and $Br(\mu\to e\gamma)$
versus  $Br(\tau\to \mu\gamma).$
The baryon asymmetry is again required to be in the range (\ref{YB}).
\vspace*{0.5cm}}
\label{fig2}
\end{figure}

To study implications of our neutrino mixing scenario and leptogenesis for
low-energy observables, we plot in Fig.~\ref{fig2} (a) the lightest
neutrino mass $m_{\nu_1}$ versus the $\beta\beta_{0\nu}$ parameter
$m_{ee}.$ There is an upper bound on $m_{\nu_1}$ coming from the
requirement of successful leptogenesis. This is an artifact of our lower
bound on the degeneracy parameter $\Delta$ and is not physical. If we
allow smaller $\Delta,$ $\epsilon_{1,2}$ are enhanced and the neutrino
mass $m_{\nu_1}$ can be higher. At high values of $m_{\nu_1},$ there can
be cancellations in $m_{ee}$ \rfn{mee}, because the phases $\phi_{2,3}$ are
free parameters not correlated with leptogenesis. However, for small
$m_{\nu_1}$ there is the definite prediction $m_{ee}= \sqrt{\Delta
m^2_{sol}} \sin^2\theta_{12},$ which is below the sensitivity of the
currently proposed $\beta\beta_{0\nu}$ decay 
experiments~\cite{bbfuture}~\footnote{If $\theta_{13}$ is not
exactly vanishing, there is an additional contribution to \rfn{mee}
and the prediction for $m_{ee}$ becomes less certain~\cite{bbth}.}.

To study LFV in our scenario, we fix the soft supersymmetry-breaking
parameters at the GUT scale to coincide with one of the post-LEP benchmark
points~\cite{bench}: $m_{1/2}=300$~GeV, $m_0=100$~GeV, $A_0=0$,
$\tan\beta=10$ and $sign(\mu)=+1.$ We plot in Fig.~\ref{fig2} (b) the
renormalization-induced branching ratio of the decay $\mu\to e \gamma$
versus the branching ratio of $\tau\to \mu \gamma.$ Because we require
$M_{N_1}$ to be relatively low, the Yukawa couplings $Y_\nu$ related to
first two families are also very small, suppressing $Br(\mu\to e \gamma)$
below the presently observable level.  Since the branching ratios scale as
$\tan^2\beta,$ reaching the level $10^{-14}$ or $10^{-15}$ planned to be
achieved at PSI~\cite{Barkov} and the neutrino factory
experiments~\cite{nufact}, respectively, is unlikely for $M_{N_1}\sim
M_{N_2}\lsim 10^8$ GeV. This indicates that the solution to the gravitino
problem and the non-observation of $\mu\to e \gamma$ in present
experiments may be related. However, $Br(\tau\to \mu \gamma)$ is not
necessarily suppressed, because $N_3$ can be heavy and the related Yukawa
couplings large. Therefore $\tau\to \mu \gamma$ may be observable in
B-factory or LHC experiments, which should achieve a sensitivity
$Br(\tau\to \mu \gamma)\sim 10^{-8}$ to $10^{-9}$~\cite{ohshima}.

In conclusion: we have studied a scenario in which the large leptonic
mixing angles $\theta_{12}$ and $\theta_{23}$ originate from $2\times 2$
mixings in the neutrino and the charged lepton Yukawa matrices,
respectively. Using a convenient phenomenological parametrization and
scanning over the all free seesaw parameters, and motivated by the
gravitino problem, we have focussed on lowering the thermal leptogenesis
scale by postulating moderate degeneracy of two lightest heavy neutrinos.
This scenario turns out to be quite predictive for $\beta\beta_{0\nu}$
decay and for the LFV decays of charged leptons, since it involves less
free parameters than the general seesaw model. We predict suppressed
$\mu\to e \gamma$, $m_{ee}= \sqrt{\Delta m^2_{sol}} \sin^2\theta_{12}$ for
hierarchical neutrinos and, in general, an observable rate for $\tau\to
\mu \gamma.$

\vskip 0.5in
\vbox{
\noindent{ {\bf Acknowledgements} } \\
\noindent  
We thank C. Gonzalez-Garcia for discussions.
This work is partially supported by EU TMR
contract No.  HPMF-CT-2000-00460, by ESF grant No. 5135,
and by Grant-in-Aid for Scientific Research (S) 14102004 (T.Y.).
}

\end{document}